\documentclass{article}

\usepackage{arxiv}

\usepackage[utf8]{inputenc} 
\usepackage[T1]{fontenc}    
\usepackage[colorlinks=true, linkcolor=blue]{hyperref}       
\usepackage{url}            
\usepackage{booktabs}       
\usepackage{amsfonts}       
\usepackage{nicefrac}       
\usepackage{microtype}      
\usepackage{lipsum}	    	
\usepackage[url=false, isbn=false, eprint=false, sorting=none, citestyle=numeric-comp, bibstyle=ieee]{biblatex}

\usepackage{doi}
\usepackage{subcaption}
\usepackage{overpic}
\usepackage{graphicx}
\usepackage{amsmath}
\usepackage{amssymb}

\usepackage{xcolor}

\usepackage{float}

\title{Compact extra dimensions as the source of\\primordial black holes}

\author{{\hspace{1mm}Valery V.~Nikulin}\textsuperscript{1}\\
\href{mailto:n-valer@yandex.ru}{\color{black}{n-valer@yandex.ru}}
\And
	{\hspace{1mm}Maxim A.~Krasnov}\textsuperscript{1}\\
	\href{mailto:morrowindman1@mail.ru}{\color{black}{morrowindman1@mail.ru}}
\And
	{\hspace{1mm}Sergey G.~Rubin}\textsuperscript{1,2}\\
	\href{mailto:sergeirubin@list.ru}{\color{black}{sergeirubin@list.ru}}
}

\date{
{\textsuperscript{1} National Research Nuclear University MEPhI (Moscow Engineering Physics
Institute),\\115409, Kashirskoe shosse 31, Moscow, Russia}\\\vspace{10pt}
{\textsuperscript{2} N. I. Lobachevsky Institute of Mathematics and Mechanics, Kazan Federal
University,\\420008, Kremlevskaya  street  18, Kazan, Russia}}

\renewcommand{\headeright}{Compact extra dimensions as the source of primordial black holes}
\renewcommand{\undertitle}{}
\renewcommand{\shorttitle}{}

\begin{document}
\maketitle

\begin{abstract}
	{We discuss the model of the primordial black holes formation at the reheating stage. These small massive black holes appear due to specific properties of the compact extra dimensions. The latter gives rise the low energy model containing the effective scalar field potential capable for the  domain walls production.  Formed during inflation, these walls are quite dense so they collapse soon after inflation ends. The discussion is performed within the scope of multidimensional $f(R)$-gravity.}
\end{abstract}


\section{Introduction}
Extra dimensions are usually studied within the framework of elementary particle physics \cite{ExtraDimInPartPhys}, for example in the context of the unification of interactions \cite{dienes1999grand,hall2001gauge}, explaining the nature of the Standard Model fields \cite{HiggsAndExtra}, searching for their manifestations in collider experiments \cite{Probes,Deutschmann2017CurrentDimensions}. This paper explores another possible cosmological consequence ---~shows that compact extra dimensions may be the cause of the primordial black holes (PBHs) formation immediately after the end of the inflation.

It is known that one of the central tasks of theories with compact extra dimensions is to provide their compactification and stabilization \cite{witten1982instability} during cosmological evolution. This can be done, for example, by introducing additional scalar fields \cite{Carroll2002} or $f(R)$-modification of gravity \cite{2007Rador, 2006PhRvD..73l4019B}. The latter approach is particularly promising because the Starobinsky quadratic $f(R)$-gravity \cite{Starobinsky:1980te, vilenkin1985classical} gives the best fit to observational constraints on the parameters of the inflation \cite{akrami2018planck}. Moreover, in multidimensional $f(R)$-gravity, the processes of cosmological inflation and compactification are manifestations of general gravitational dynamics in different subspaces \cite{fabris2020multidimensional}.

The possibilities of $f(R)$-gravity are widely studied \cite{fR, ExtendedFR}, they offer solutions to many cosmological problems \cite{fRtoDM, Bronnikov_2007, Bronnikov_2017, InhomDims}. One of the problems that $f(R)$-gravity can solve is the existence of primordial black holes. Today, the primordial origin of some discovered black holes (quasars at small $z$ \cite{10.1093/mnras/stu283,dokuchaev2007origin}, BHs of intermediate masses detected by gravitational-wave observatories \cite{InterPBH}) is hotly discussed \cite{carr2021primordial, LIGOVirgo, Sakharov:2021dim}.
In this paper we demonstrate how the primordial black holes can appear as a result of inflationary dynamics in the framework of $f(R)$-gravity model.

The idea of our proposed mechanism is based on the known possibility of domain walls formation during cosmological inflation followed by their collapse into primordial black holes \cite{PBHfromtransition,Clusters}.
The formation of such domain walls requires a scalar field with a nontrivial potential containing several vacuums. Such kind of scalar field effectively arises in the multidimensional $f(R)$-models in the Einstein frame \cite{lyakhova2018classical,2006PhRvD..73l4019B,fabris2020multidimensional}. This field controls the size of the compact extra space, and its different vacuums correspond to different universes. In this paper we calculate the parameters of the domain walls formed by the field and conclude that appearing at inflationary stage they will immediately collapse into PBHs during reheating. For a remote observer in the Jordan frame the appearance of such PBH is interpreted as a manifestation of non-trivial $f(R)$-gravitational dynamics of multidimensional space. These formed PBHs will grow rapidly in the process of further cosmological evolution due to accretion and are capable to turn into observable supermassive quasars at small $z$.

\section{Model}

The study is based on the modified $f(R)$  gravity acting in $D=4+n$ dimensions. It is described by the action\footnote{
In this paper we use the following conventions for the Riemann curvature tensor $R_{\mu\nu \alpha}^{\beta}=\partial_{\alpha}\Gamma_{\mu \nu}^{\beta}- \partial_{\nu}\Gamma_{\mu \alpha}^{\beta}+\Gamma_{\sigma \alpha}^{\beta}\Gamma_{\nu\mu}^{\sigma}-\Gamma_{\sigma \nu}^{\beta}\Gamma_{\mu \alpha}^{\sigma}$, and the Ricci tensor is defined as follows: $R_{\mu \nu}=R^{\alpha}_{\mu \alpha \nu }\,.$}
\begin{equation}\label{act0}
S =\frac{m_{D}^{D-2}}{2} \int d^{4+n} x \sqrt{|g_D|}\, \left[f(R) +c_1 R_{AB} R^{AB}+c_2 R_{ABCD} R^{ABCD}\right]\,,\nonumber
\end{equation}
\begin{equation}\label{fR1}
f(R)=a_2 R^2 + R - 2\Lambda_D\,,
\end{equation}
where $m_D$ is the multidimensional Planck mass. {The parameters $a_2, c_1, c_2$ of the Lagrangian \eqref{act0} have dimensionality $[m_D^{-2}]$, the cosmological constant $\Lambda$ has dimensionality $[m_D^2]$. Hereafter we will work everywhere in units $m_D\equiv1$, unless it is explicitly specified.} In this work we will represent multidimensional space as the direct product of $\mathbb{M} = \mathbb{M}_4 \times \mathbb{M}_n$, where $\mathbb{M}_4$ is-- four-dimensional space, $\mathbb{M}_n$ -- an $n$-dim compact extra space with metric
\begin{equation}
    ds^2 = g_{\mu \nu} dx^\mu dx^\nu - e^{2\beta(t,x)} d\Omega^2_n\,.
\end{equation}
Here $g_{\mu \nu}$ is a 4-dimensional metric $\mathbb{M}_4$, $\beta$ is a metric function and $d\Omega^2_n$ is the linear element square of a maximally symmetric compact extra space $\mathbb{M}_n$ with positive curvature.

Cosmological scenarios based on this theory \cite{lyakhova2018classical, fabris2020multidimensional, bronnikov2021local}. are finished by an effective 4-dim theory at the low energy. Its properties are defined by the Lagrangian  parameters of of action \eqref{act0}. The procedure for obtaining such a theory is also described in \cite{2006PhRvD..73l4019B}.

Following this procedure, we assume maximal symmetry of the extra space $\mathbb{M}_n$. Its radius $\rho\equiv e^\beta$ does not depend on the extra coordinates. In addition, the useful approximation of this effective theory works in a region where the extra curvature $R_n = n(n-1)/e^{2\beta}$ is large compared to the 4-dimensional curvature and changes slowly:
\begin{equation}
R = R_4 + R_n + P_k\,,\quad P_k = 2 n\,\partial^2 \beta + n(n+1) (\partial \beta)^2\,,\nonumber
\end{equation}
\begin{equation}\label{effective}
R_4, P_k \ll R_n\,.
\end{equation}
where $R_4, R_n$ are Ricci scalars for $\mathbb{M}_4, \mathbb{M}_n$. The approximation \eqref{effective} applied to action \eqref{act0} leads to the effective 4-dim model with a non-minimal coupling between the observed 4-dimensional gravity $R_4$ and the scalar field \cite{2006PhRvD..73l4019B}. Therefore the resulting theory will be written in the Jordan frame and we will use this frame as the physically observed one \cite{PhysNotEq}.


\section{Effective scalar field action}

The minimal coupling between the observed 4-dimensional gravity and the scalar field is achieved by the transition to the Einstein frame, which strongly facilitate the analysis. This transition has been performed in detail in \cite{2006PhRvD..73l4019B}, \cite{2013bhce.conf.....B}:
\begin{equation}\label{act_eff}
S = \frac{m_4^2}{2}\int d^4x \sqrt{-g_4}\  \text{sign}(f') \left[R_4 + K(\phi) (\partial \phi)^2 - 2 V(\phi) \right]\,,
\end{equation}
where the Planck mass in the Einstein frame is $m_4 = \sqrt{2\pi^{\frac{n+1}{2}} / \Gamma(\frac{n+1}{2})}$, and $g_4^{\mu\nu}$ is the observed 4-dimensional metric. The Ricci scalar of the extra space is perceived as the scalar field, $\phi \equiv R_n$.

The action \eqref{act_eff} contains a potential and a nontrivial kinetic term, which are expressed through the initial parameters of the Lagrangian \eqref{act0} (see \cite{fabris2020multidimensional}):
\begin{equation}\label{Kin}
K(\phi)=\frac{1}{4 \phi^{2}}\left[6 \phi^{2}\left(\frac{f^{\prime \prime}}{f^{\prime}}\right)^{2}-2 n \phi\left(\frac{f^{\prime \prime}}{f^{\prime}}\right)+\frac{n(n+2)}{2}\right]+\frac{c_{1}+c_{2}}{f^{\prime} \phi}\,,
\end{equation}
\begin{equation}\label{Pot}
V(\phi)=-\frac{\operatorname{sign}\left(f^{\prime}\right)}{2 (f^{\prime})^2}\left[\frac{|\phi|}{n(n-1)}\right]^{n / 2}\left[f(\phi)+\frac{c_1+2c_2/(n-1)}{n} \phi^{2}\right]\,.
\end{equation}
The transition validity from \eqref{act0} to \eqref{act_eff} dictated by ineaqualities \eqref{effective} will be analyzed later.

\begin{figure}[H]
\begin{center}
    \includegraphics[width=12cm]{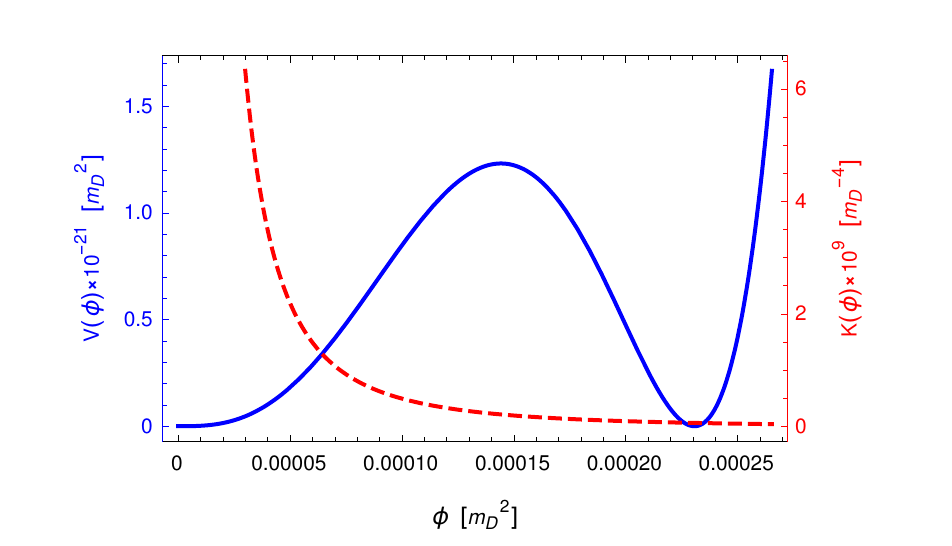}
    \caption{Graphs of the potential term and kinetic factor \eqref{Kin}, \eqref{Pot} for parameters: 
    $n=6$, $c_1=-8000$, $c_2=-5000$, $a_2=-500$.
    The left minimum of the potential is at $\phi=0$, but it is impossible to roll down to it in a finite time because of the increasing friction effect of the kinetic factor, the right minimum is at $\phi_{min}=2.3\cdot 10^{-4}.$}
   \label{potential&kinetic}
   \end{center}
\end{figure}

For a wide range of parameters, the potential term \eqref{Pot} has two minima corresponding to different vacuums of the Universe (Fig.~\ref{potential&kinetic}). The value of the potential at these minima must be almost zero according to almost zero value of the cosmological constant $\Lambda_4$. It leads to a relation of the parameter $c =1/4(a + c_v/n)$, where $c_v = c_1 + 2c_2/(n-1)$.

The rolling of the field $\phi\equiv R_n$ into the right minimum of the potential \eqref{Pot} corresponds to the stabilization of the compact extra space $R_n\neq 0$ (the extra space is compactified and has some radius $\rho_0$), and leads to the observed cosmology \cite{fabris2020multidimensional}. The presence of the left minimum $R_n\approx0$ suggests the possibility of another scenario \cite{bronnikov2021local}, in which the extra space is unstable and expands to macroscopic scale. 

The non-trivial kinetic factor \eqref{Kin} significantly modifies the character of the field evolution $\phi(t)$ in comparison to the standard scalar-field theory, providing an increasing friction when rolling to the left minimum, see Fig.~\ref{potential&kinetic}. One can simplify the Lagrangian by substituting
\begin{equation}\label{change}
\psi = m_4\int\limits_{\phi_0}^{\phi} \sqrt{K(\phi')}\,d\phi'\,,\quad  \tilde{V}(\psi) = m_4^2\, V(\phi(\psi)), \quad K(\phi')>0\,.
\end{equation}
In this case, $\psi(\phi)$ is monotonic and invertible (which is required to find the potential in the expression \eqref{act2}).
After simple algebra, the Lagrangian is reduced to the standard form
\begin{equation}\label{act2}
S = \frac{m_4^2}{2}\int d^4x \sqrt{-g_4}\, R_4 + \int d^4x \sqrt{-g_4}\,\left[\frac{1}{2}(\partial \psi)^2 - \tilde{V}(\psi) \right]\,,
\end{equation}
where $m_4^2=16 \pi ^{3}/{15}$ for chosen extra space dimensionality $n=6$.

\begin{figure}[ht]
\begin{center}
    \includegraphics[width=12cm]{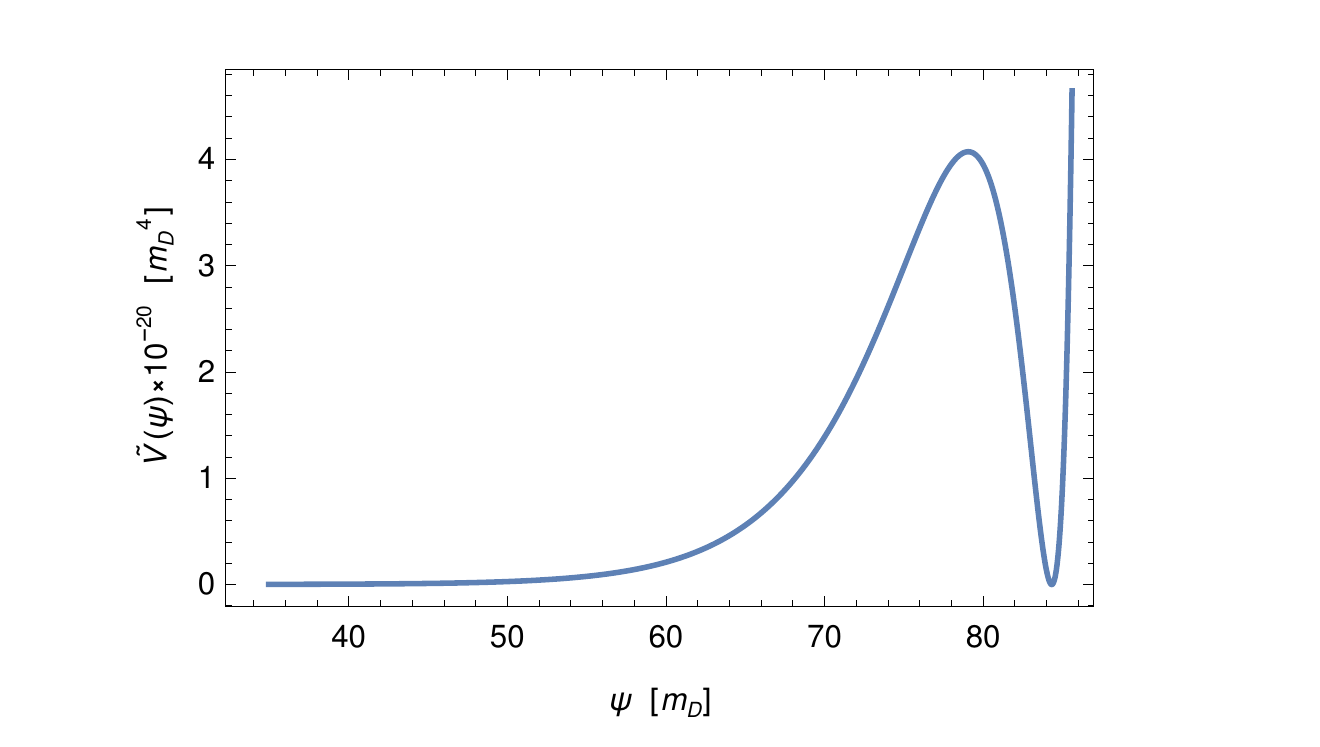}
    \caption{Graph of the potential term $\tilde{V}(\psi)$ of the effective theory \eqref{act2} for the parameters chosen in Fig.~\ref{potential&kinetic}. The right minimum is at $\psi_{min}=84.3$. Strictly speaking, the left minimum is at $\psi=-\infty$, but for our purposes a very accurate approximation $\psi=0$ is sufficient (corresponding to $\phi=\phi_0$), up to which the effective theory \eqref{act_eff} still works. The maximum value of the potential in the transition region between vacuums $\tilde{V}_{max}\approx4.1\cdot10^{-20}$ and the steepness of the slope $\sqrt{\tilde{V}''(\psi_{max})} = 2.7\cdot10^{-11}$.}
\label{pot_psi}
\end{center}
\end{figure}

\section{Domain walls}

It is well known that potentials like \eqref{pot_psi}, containing several minima (vacuums), can lead to formation of non-trivial field configurations \cite{vilenkin1994cosmic} --- "bubbles"\ of one vacuum inside another, surrounded by a domain wall.

{Let us investigate such a configuration by deriving the field equation for $\psi$ from the effective action \eqref{act2}. For simplicity, we consider it spherically symmetric and static, which gives the equation:
\begin{equation}\label{field_eq}
    \psi_{uu} + \frac{2\psi_{u}}{u} - \tilde{V}'(\psi) = 0\,,
\end{equation}
where $u$ is the radial coordinate.} When considering a sufficiently large "bubble"\ (such that its radius is much larger than the characteristic thickness of the domain wall $\psi_{uu} \gg 2\psi_{u}/u$) it can easily be reduced to a first-order equation
\begin{equation}\label{field_eq2}
    \partial_u\psi = \pm\sqrt{2\tilde{V}(\psi)}\,.
\end{equation}

\begin{figure}[ht]
    \centering
    \includegraphics[width=12cm]{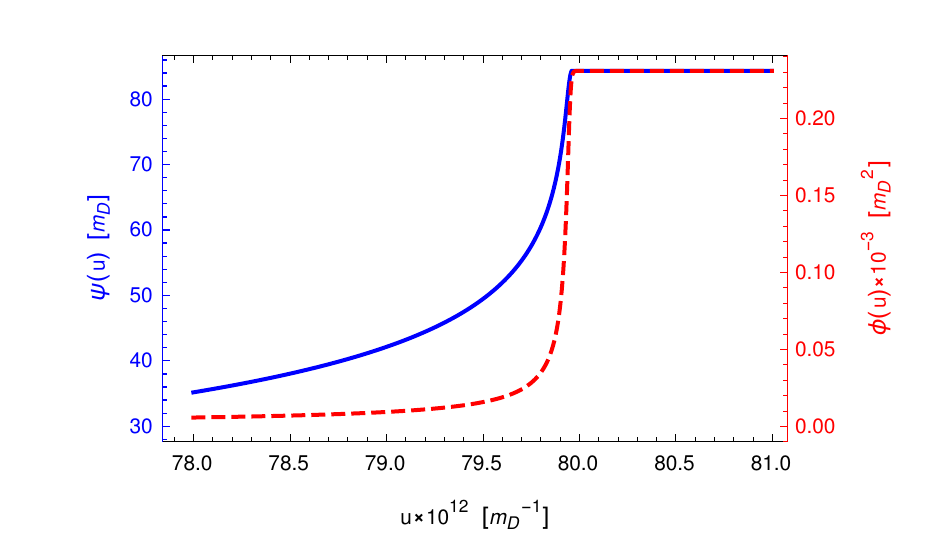}
    \caption{Numerical solution of \eqref{field_eq} under boundary conditions: the left minimum $\tilde{V}(\psi)$ (Fig.~\ref{pot_psi}) is reached inside the bubble $\psi(0)=0$, and the right minimum is reached for the remote observer $\psi(u\rightarrow\infty)=\psi_{min}$ (and forms our Universe). In the graphs: the blue line is the direct solution of $\psi(u)$ (domain wall) and the red dashed line is its corresponding function $\phi(u)$. In this example the bubble radius is $u_0 \approx 8\cdot 10^{13}$.}
    \label{walls}
\end{figure}

The characteristic solution of the equation \eqref{field_eq2} connecting the left vacuum of the potential $\tilde{V}(\psi)$ to the right one (Fig.~\ref{pot_psi}) is shown in Fig.~\ref{walls} (blue line). 

We can calculate the domain wall energy density $\varepsilon_\psi$ as a component  $T^{00}$ of energy-momentum tensor for the scalar field Lagrangian $\psi$:
\begin{equation}\label{T00}
    \varepsilon_\psi(u) = T^{00}(u)=\cfrac{\partial \mathcal{L}_\psi}{\partial \left(\partial_{0} \psi \right)}\,\partial^{0} \psi - \mathcal{L}_\psi g^{00} = \frac{1}{2}(\partial_u \psi)^2 + V(\psi) = 2 \tilde{V}(\psi(u)).
\end{equation}
Integration over the radial coordinate \eqref{T00} gives the surface energy density of the domain wall in multidimensional Planck units ($m_D = 1$):
\begin{equation}\label{surface}
    \sigma=\int \limits_0^\infty \varepsilon_\psi(u)\,du = \int \limits_0^{\psi_{min}} \frac{2 \tilde{V}(\psi)}{\psi_u}\,d\psi = \int \limits_0^{\psi_{min}} \sqrt{2 \tilde{V}(\psi)}\,d\psi\,.
\end{equation}
We can also formally estimate the characteristic wall thickness $\delta$ as
\begin{align}\label{thickness}
    \sigma = & \int \limits_0^\infty 2 \tilde{V}(\psi)\,du \approx 2\frac{\tilde{V}_{max}}{2}\,\delta \quad\implies \quad \delta \approx \frac{\sigma}{\tilde{V}_{max}}\,.
\end{align}

\begin{figure}[ht]
\begin{center}
    \includegraphics[width=12cm]{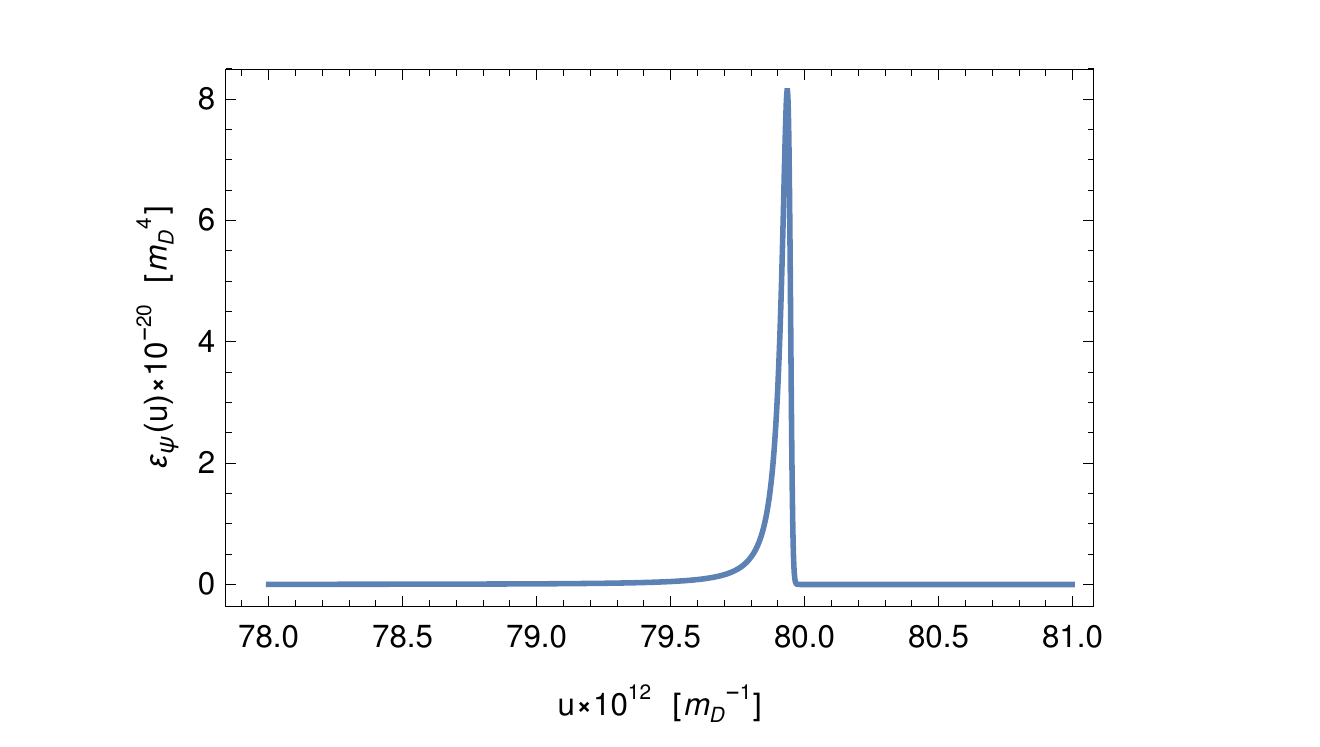}
    \caption{Energy density \eqref{T00} profile along the radial coordinate $u$ for the solution shown in Fig.~\ref{walls}. {In the case of spherical symmetry, the domain wall forms a bubble with surface energy density $\sigma \approx 5\cdot10^{-9}$ and wall thickness $\delta \approx 1.2\cdot10^{11}$, according to \eqref{surface}, \eqref{thickness}.} } 
   \label{energydensity}
\end{center}
\end{figure}

Domain walls considered in this paper appears to be very massive (Fig.~\ref{energydensity}). As a result, they could be a source of primordial black holes. The mechanism of formation and collapse of such domain walls is well studied in \cite{Liu_2020,2016, PBHfromtransition} and leads to the formation of primordial black holes in a wide range of masses.

\section{Generation of domain walls during inflation}\label{constr}

As stated above, the rolling of the $\psi$ field to the right minimum of Fig. \ref{pot_psi}, creates the observable Universe. For its formation, a mechanism of the cosmological inflation is required. In order not to complicate our consideration, we will consider the inflation as an external process with a characteristic Hubble parameter $H$.

The mechanism of inflationary production of the bubbles of alternative vacuum is well known \cite{PBHfromtransition}. As a result of repeated quantum fluctuations, the field $\psi$ can be flipped from the region of rolling down to the right minimum to the region of rolling down to the left minimum in some area of the inflationary Universe. Thy size of this region is growing during inflation while the $\psi$ field is frozen near the potential maximum. After the end of inflation, the field tends to one minimum inside the bubble and another minimum outside it. {Increasing energy density gradually forms the domain wall (Fig.~\ref{energydensity}) around the bubble.}

Several constraints must be imposed on the working model of domain wall production in the considered $f(R)$-gravity:
\begin{enumerate}
    \item\label{const1} During cosmological inflation, 4-dimensional space is described by a de Sitter metric with curvature $R_4 \simeq 12 H^2$, where $H$ is the Hubble parameter. Therefore, the approximation \eqref{effective} is applicable only for field values $\phi \equiv R_n \gg R_4 \simeq 12H^2$. {From Fig.~\ref{potential&kinetic} we see that $\phi\sim10^{-6}-10^{-4}$, so to satisfy the constraint we need $H\ll10^{-4}$.}
    
    \item\label{const2} To generate walls via quantum fluctuations of field $\psi$ (near the maximum of potential \ref{pot_psi}) its slow rolling is required: $\sqrt{\tilde{V}''(\psi_{max})}\ll H$. {According to Fig.~\ref{pot_psi} we see that $\sqrt{\tilde{V}''(\psi_{max})} \sim 10^{-11}$, so to satisfy the constraint we need $H\gg10^{-11}$.}
    
    \item\label{const3} Domain walls should not be too dense so as not to dominate the inflaton: $\varepsilon_\psi \ll \varepsilon_{inf} \sim H^2 m_4^2$. {From Fig.~\ref{energydensity} we see that $\varepsilon_\psi \sim 10^{-20}$, so to satisfy the constraint we need $H\gg10^{-11}$.}
    
    \item\label{const4} The fluctuations of the $\psi$ field during inflation should not be too large to prevent overproduction of the domain walls: $\delta\psi = H/2\pi \ll \psi$. {As one can see from Fig.~\ref{pot_psi}, $\psi \sim 10$, so to satisfy the constraint we need $H\ll10$.}
\end{enumerate}
{All constraints above are satisfied if the inflation has the characteristic scale $10^{-11} \ll H \ll 10^{-4}$. Here $m_4=5.75$ for $n=6$ according to \eqref{act_eff}. Remind that all estimations are given in the Einstein frame and if the dimensionality is not explicitly specified, all calculations are carried out in $m_D\equiv1$ units.}

In the previous sections, all calculations were performed in the Einstein frame. As a physically observable one it is often accepted to consider the Jordan frame, in which it is possible to manipulate the 4-dimensional Planck scale $M_4 = \Omega(\phi_\text{min}) m_4$ by adjusting the size of extra space. {So we will transform final physical results to the Jordan frame.} Cosmological inflation should be described in a physically observable frame --- that is the Jordan frame (let's denote it by the index $J$):
\begin{equation}\label{infJ}
S^J_\text{inf} = \int d^4x \sqrt{-g_4^J}\,\left[\frac{1}{2}\left(\partial \chi^J\right)^2 - U^J\left(\chi^J\right) \right]\,,
\end{equation}
where $\chi$ is the inflaton whose potential $U(\chi)$ determines the Hubble parameter $H\sim\sqrt{G\,U(\chi)}$ during the inflation. The constraint above on $H$ have been checked for the inflation in the Einstein frame (denote it by the index $E$). The transition to it from \eqref{infJ} is known \cite{ConfOneField,fabris2020multidimensional}:
\begin{equation}\label{conform}
g_{\mu\nu}^J = \Omega^{-2} g_{\mu\nu}^E\,,\quad \text{where} \quad \Omega^{2} = e^{n\beta(\phi)}|f'(\phi)|.
\end{equation}
Substituting \eqref{conform} into \eqref{infJ}, we obtain the action written in Einstein frame:
\begin{align}\label{infE}
S^J_\text{inf} &= \int d^4x \sqrt{-g_4^E}\, \left[\frac{1}{2}\Omega^{-2}\left(\partial \chi^J\right)^2 - \Omega^{-4}U^J\left(\chi^J\right) \right]\approx \nonumber\\
&\approx \int d^4x \sqrt{-g_4^E}\, \left[\frac{1}{2}\left(\partial \chi^E\right)^2 - U^E\left(\chi^E\right) \right] = S^E_\text{inf}\,,
\end{align}
where transformations of the inflaton $\chi^E=\Omega^{-1}\chi^J$ and its potential $U^E=\Omega^{-4}U^J$ to the Einstein frame are made.
Here we use the fact that during inflation the field $\phi$ is practically frozen (constraint \ref{const2}) and is in the region of maximum of potential, so the factor $\Omega(\phi)$ can be considered constant and equal $\Omega\approx\Omega(\phi_\text{max}) \sim \Omega(\phi_\text{min}) \sim 10^8$ for the parameters chosen in Fig.~\ref{potential&kinetic}.

The Hubble parameter for the inflaton  $\chi^E$ observed in the Einstein frame is related to the Hubble parameter for the $\chi^J$ inflaton observed in the Jordan frame as follows:
\begin{equation}\label{HEtoHJ}
H^E\sim\sqrt{G\,U^E(\chi)} = \sqrt{G\ \Omega^{-4} U^J(\chi)} \sim \Omega^{-2} H^J\,.
\end{equation}
From this, in accordance with the constraint for $H$ obtained above for the Einstein frame, it follows that $H^J > 10^5 \sim 10^{-4}\ [M_4] \sim 10^{14}\ [\text{GeV}]$ at the chosen parameters. It is consistent with the observed data \cite{akrami2018planck}.


\section{Formation of primordial black holes}
During the inflationary stage the classical motion of the scalar field $\psi$ is frozen --- this is determined by the constraint \ref{const2} in Section \ref{constr}). If there a seed for the future domain wall appear during quantum fluctuations (the field jumps to the left slope of the potential in Figure \ref{pot_psi}), then in this place the field values $\psi$ should lie near the potential maximum since the fluctuations are small (constraint \ref{const4} in Section \ref{constr}). At the end of inflation, the inflaton the Hubble parameter vanishes and the $\psi$ field  begins to rapidly roll from the maximum to the potential minima (to the left minimum, for the inner region of the bubble and to the right minimum for the outer region). During this roll-off, the energy density  $\sigma(t)$ in the transition region gradually grows up to the value $\sigma$ calculated in the previous sections. In addition, due to postinflationary expansion, the radius of the region is growing $u_w(t)=u_0 a(t)$ also, where $u_0$ is radius at the end of inflation and $a(t)$ is scale factor. Because of this, the mass of the formed domain wall increases $m_w(t) = 4\pi u_w(t)^2 \sigma(t)$ together with its gravitational radius $u_g(t)=2Gm_w(t)$. At a certain time, the gravitational radius will cross the entire domain wall $u_g(t_*)=u_w(t_*)$ and a primordial black hole with mass $m_w(t_*)$ will form for a remote observer.

The moment $t_*$ of crossing of the wall by the gravitational radius in our model will come long before it reaches the final energy density $\sigma$. This can be seen from the fact that the ratio of the gravitational radius to the size of the wall (assuming that the wall has a final energy density $\sigma$) is grater than unity
\begin{equation}\label{critRad}
\frac{u_g}{u_w} = 8\pi G \sigma u_w > m_4^{-2} \sigma \delta \approx 16\,,
\end{equation}
where the fact that the wall radius is always larger than its thickness $u_w(t)>\delta$ is used. The ratio \eqref{critRad} is calculated for the wall parameters obtained in Fig.~\ref{energydensity} and holds for any model parameters \eqref{fR1} {satisfying the set of constraints discussed in Section \ref{constr}.} 
Therefore, the wall is formed long before the field rolls to the potential minima. The black hole is formed at the time $t_*$ when the horizon radius $u_g$ is approximately equal to the wall radius $u_w$.  Hence, the mass of the PBH can be estimated as
\begin{equation}\label{MPBH}
M_{PBH} \simeq m_w(t_*) \simeq u_g(t_*)/2G \simeq u_w(t_*)/2G  \equiv u_0 a(t_*)/2G\,.
\end{equation}
All processes described above occur in a very short time interval $t_*$, which is significantly less than the characteristic time of the field $\psi$ roll-off to the minimum of potential, $t_*<\tau_{\psi}\sim \sqrt{V''(\psi_{\text{max}})}^{-1}\sim 10^{6}$. This time in our model is less than the inflation time $\tau_{\text{inf}} \approx 60\ H^{-1}\sim 10^{7}$. Therefore, the roll-off of field $\psi$ process occurs at the reheating stage (as for the inflaton), the scale factor at which we can approximately \cite{lazarides2002inflationary} consider $a(t_*)=\left((\tau_{\text{inf}}+t_*)/\tau_{\text{inf}}\right)^{\frac{\nu+2}{3\nu}}<\left((\tau_{\text{inf}}+\tau_{\psi})/\tau_{\text{inf}}\right)^{\frac{\nu+2}{3\nu}}\sim\left(1+10^{- 1}\right)^{\frac{\nu+2}{3\nu}}\sim1$, where $\nu$ is the index of degree in the inflaton potential. It means that the time interval $t_*$ is very short and one can neglect the post-inflationary expansion.

The radius $u_0$ is determined by the moment of inflationary generation of the fluctuation, leading to the subsequent formation of the wall. If the fluctuation is formed at the e-fold $N$ of total number $N_{\text{inf}}$, then its size by the end of inflation will be $u_0 = H_{\text{inf}}^{-1}e^{N_\text{inf} - N}$. {For example, suppose that the fluctuation of the $\psi$ field leading to the wall formation occurred at $N=25$ e-fold of $N_\text{inf}=60$ and the inflation scale is $H\sim 10^{14}$~GeV (which are characteristic parameters \cite{akrami2018planck}) Then, immediately after the inflation ends, a PBH of mass $M_{PBH}\sim  u_w/2G  \sim u_0/2G = 4\pi M_4^2 H_{\text{inf}}^{-1}e^{35} \sim 10^{41}\ [GeV] \sim 10^{-16}\ [M_\odot]$  is formed. During further cosmological evolution in the reheating stage, such a PBH will gain mass as a result of accretion. The dynamics of this process is rather complicated and depends on the properties of the reheating stage \cite{carr2018primordial}.} Since such black holes are formed at the super-dense Universe, immediately after the end of inflation, their masses can reach many solar masses due to the accretion, as our estimates show.

{Calculation of the initial mass spectrum $N(M)$ of the above described PBHs is reduced to the calculation of the size spectrum $N(u)$ for the $\psi$ field fluctuations generated at the inflationary stage. It can be done by substituting the connection $u \equiv u_w = 2G M$.} The spectrum $N(u)$ is calculated in such works as \cite{Clusters, nikulin2017mechanism, FormationFromPhase}. They investigated the dependence of this spectrum on inflation parameters and the initial field value $\psi_0$, with which the modern horizon appears. In these works it was shown that the width of the spectrum, the characteristic masses, and the total number of PBHs strongly depend on the choice of $\psi_0$. We will not investigate the spectrum in the present paper, since its goal was to demonstrate a new mechanism for PBH formation, arising in $f(R)$-gravity with the extra dimensions.

\section{Conclusion}
In this paper we discuss the model of multidimensional $f(R)$-gravity that gives rise the primordial black holes formation at the early stage of cosmological evolution. For a remote observer in the Jordan frame, the formation of such black holes can be interpreted as a manifestation of non-trivial gravitational dynamics of multidimensional space after the cosmological inflation.

In this paper the dynamics of the extra dimensional metric is reduced to the dynamics of the effective scalar field. The domain walls formation occurs just after the completion of cosmological inflation. We show that PBH nucleation is inevitable in our model for that range of parameters of $f(R)$-gravity which both satisfy cosmological constraints and lead to a non-trivial effective scalar field potential. The mechanism of collapse into primordial black holes has been studied in \cite{Clusters, nikulin2017mechanism, FormationFromPhase}.

{Primordial black holes arising in the developed mechanism appear at the reheating stage and then will actively grow due to accretion. In addition, the formation of several domain walls under one gravitational radius is possible, which modifies our mechanism. These rather complex processes will affect the formation of the final mass spectrum of primordial black holes and will be considered by us in future works.}

{Cosmological consequences of multidimensional $f(R)$-gravity are quite rich and lead to a variety of exotic observational manifestations. As our study shows, the widely discussed primordial black holes are added to such manifestations. Therefore, the confirmation of the primordial origin of some classes of black holes may be the evidence in favor of existence of extra dimensions.}

\section{Acknowledgements}

The work of V.V.N. and M.A.K. was supported by the MEPhI Program Priority 2030.
The work of S.G.R. has been supported by the Kazan Federal University Strategic Academic Leadership Program.

\printbibliography

\end{document}